# Asynchronous Ripple Carry Adder based on Area Optimized Early Output Dual-Bit Full Adder

P. Balasubramanian
School of Computer Science and Engineering
Nanyang Technological University
50 Nanyang Avenue
Singapore 639798

## Abstract

This technical note presents the design of a new area optimized asynchronous early output dual-bit full adder (DBFA). An asynchronous ripple carry adder (RCA) is constructed based on the new asynchronous DBFAs and existing asynchronous early output single-bit full adders (SBFAs). The asynchronous DBFAs and SBFAs incorporate redundant logic and are encoded using the delay-insensitive dual-rail code (i.e. homogeneous data encoding) and follow a 4-phase return-to-zero handshaking. Compared to the previous asynchronous RCAs involving DBFAs and SBFAs, which are based on homogeneous or heterogeneous delay-insensitive data encodings and which correspond to different timing models, the early output asynchronous RCA incorporating the proposed DBFAs and/or SBFAs is found to result in reduced area for the dual-operand addition operation and feature significantly less latency than the asynchronous RCAs which consist of only SBFAs. The proposed asynchronous DBFA requires 28.6% less silicon than the previously reported asynchronous DBFA. For a 32-bit asynchronous RCA, utilizing 2 stages of SBFAs in the least significant positions and 15 stages of DBFAs in the more significant positions leads to optimization in the latency. The new early output 32-bit asynchronous RCA containing DBFAs and SBFAs reports the following optimizations in design metrics over its counterparts: i) 18.8% reduction in area than a previously reported 32-bit early output asynchronous RCA which also has 15 stages of DBFAs and 2 stages of SBFAs, ii) 29.4% reduction in latency than a 32-bit early output asynchronous RCA containing only SBFAs.

## 1. Introduction

The conventional full adder (i.e. SBFA) is used to add two binary inputs (i.e. a 1-bit augend and a 1-bit addend) along with a 1-bit carry input from a preceding stage and produces two 1-bit binary outputs viz. a 1-bit sum and a 1-bit carry output. The full adder can be realized in synchronous [1] – [4] or asynchronous design style [5] – [15]. As an alternative to the conventional SBFA, the DBFA was proposed in [16] – [18], [43] based on the synchronous and asynchronous design paradigms. The DBFA adds a 2-bit augend with a 2-bit addend binary input along with a 1-bit carry input from a preceding stage and produces 2-bit sum outputs and a 1-bit carry output. It was shown in [16] – [18], [43] that irrespective of whether the design style adopted is synchronous or asynchronous, the DBFA when cascaded to form a RCA would substantially reduce the (forward) latency, i.e. the critical path delay for processing the valid data of a RCA that is constructed using only SBFAs.

In this work, we discuss the design of an early output asynchronous RCA that incorporates DBFAs and SBFAs which are based on the delay-insensitive dual-rail data encoding scheme (i.e. homogeneous data encoding) and corresponds to the 4-phase return-to-zero handshaking. The proposed asynchronous RCA involving DBFAs and SBFAs when used to perform 32-bit addition results in reduced latency and area compared to its counterpart RCAs comprising only SBFAs or DBFAs which are based on homogeneous or heterogeneous delay-insensitive data encoding. This inference is based on simulation results obtained by using a 32/28nm CMOS process.

The remainder of this work is organized into five sections. A background about asynchronous circuit design that uses delay-insensitive data encoding (homogeneous or heterogeneous), and which adheres to the 4-phase (return-to-zero) handshake protocol is provided in Section 2. The proposed asynchronous DBFA design and the asynchronous RCA involving DBFAs and SBFAs is discussed in Section 3. Next, the simulation results corresponding to various 32-bit asynchronous RCAs incorporating DBFAs and/or SBFAs are presented and compared in Section 4. Finally, Section 5 concludes this work.

## 2. Robust Asynchronous Circuits – Design Fundamentals
### 2.1. Asynchronous Circuit Stage Operation

An asynchronous function block is the equivalent of the synchronous combinational logic [19]. When an asynchronous function block is constructed using delay-insensitive codes [20] and utilizes a 4-phase (return-to-zero) handshaking, it is generally robust provided it is free of gate and wire orphans [21] – [23]. Orphans are unacknowledged signal transitions which may occur on gate outputs (gate orphans) or in wires (wire orphans). Wire orphans are eliminated by imposing the isochronicity assumption [24], which forms the weakest compromise to delay-insensitivity. An isochronic fork implies that a signal transition on a wire junction i.e. a node is concurrently transmitted across all the wire branches. Gate orphans may however become problematic and so their possibility of occurrence should be eliminated to guarantee that an asynchronous circuit based on delay-insensitive data encoding and 4-phase handshaking would be robust.

The dual-rail code, also called the 1-of-2 code, is the simplest member of the family of delay-insensitive $m$-of-$n$ data codes [20]. Among the family of $m$-of-$n$ codes, the 1-of-$n$ codes represent a subset and are called one-hot codes. In a 1-of-$n$ code, only 1 out of $n$ wires is asserted high, i.e. binary 1 to represent a binary data. In fact, the 1-of-$n$ coding scheme is said to be unordered [25] since none of the code words forms a subset of any other code word. Further, the 1-of-$n$ coding scheme is said to be complete [26] if all the $n$ unique code words as per the definition are utilized to encode the specified binary data. Table 1 shows an example binary data represented using the 1-of-2 and 1-of-4 data encoding schemes.

Table 1. 2-bit binary data represented using 1-of-2 and 1-of-4 data encoding schemes

| Binary data | | 1-of-2 encoded data | | 1-of-4 encoded data | | | |
|---|---|---|---|---|---|---|---|
| P | Q | (P1, P0) | (Q1, Q0) | F0 | F1 | F2 | F3 |
| 0 | 0 | (0, 1) | (0, 1) | 1 | 0 | 0 | 0 |
| 0 | 1 | (0, 1) | (1, 0) | 0 | 1 | 0 | 0 |
| 1 | 0 | (1, 0) | (0, 1) | 0 | 0 | 1 | 0 |
| 1 | 1 | (1, 0) | (1, 0) | 0 | 0 | 0 | 1 |

As per the 1-of-2 code, a single-rail binary input, say X, is encoded using two wires as say X1 and X0, where X = 1 is represented by X1 = 1 and X0 = 0, and X = 0 is represented by X1 = 0 and X0 = 1. X1 and X0 cannot simultaneously assume 1 as it is illegal and invalid because the coding scheme will not remain unordered. However, X1 and X0 can assume 0 simultaneously and this is referred to as the spacer. Hence as per the 1-of-2 code, a valid data is specified by either X1 or X0 assuming binary 0 and the other assuming binary 1, and the condition of both X1 and X0 assuming binary 0 is labelled the spacer or null i.e. not data. On the other hand, the 1-of-4 code is used to represent two bits of binary information at a time. Referring to Table 1, the two binary inputs specified by P and Q are encoded into F0, F1, F2 and F3 as per the 1-of-4 code for an example. When just one delay-insensitive code (say, 1-of-2 code) is alone used to encode the given binary data, it is called homogeneous delay-insensitive data encoding, and when more than one delay-insensitive code (for example, 1-of-2 and 1-of-4 codes) is used to encode the specified binary data, it is called heterogeneous delay-insensitive data encoding.

A typical asynchronous circuit stage that employs delay-insensitive codes for data encoding and processing, and the 4-phase return-to-zero handshake protocol for data communication is shown in Fig. 1. As the name suggests, the 4-phase return-to-zero handshake protocol consists of 4 phases. This will be explained with reference to Fig. 1 based on the assumption that the dual-rail code is used for data representation. Nevertheless, the explanation would remain applicable for data represented using any delay-insensitive 1-of-$n$ code.

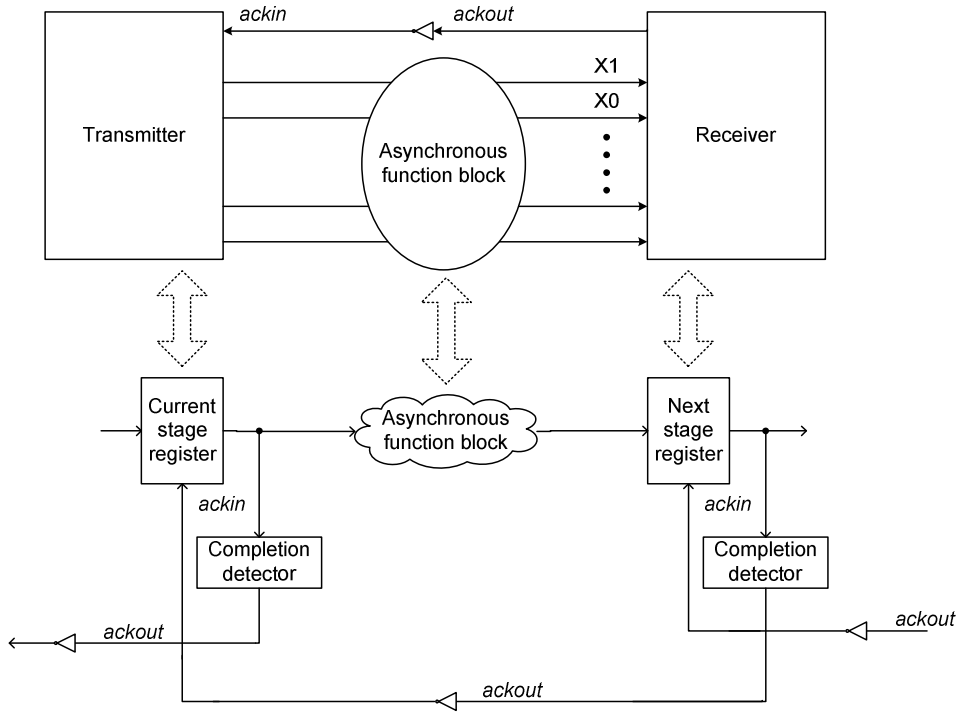

Fig. 1. A robust asynchronous circuit stage operation that is correlated with the transmitter-receiver analogy

In the first phase, the dual-rail data bus shown in Fig. 1 is in the spacer state and *ackin* is high i.e. binary 1. The transmitter now transmits a code word i.e. valid data and this results in upgoing signal transitions on any one of the corresponding dual rails of the entire dual-rail data bus. In the second phase, the receiver receives the code word sent, and it drives *ackout* high. In the next phase viz. the third phase, the transmitter waits for *ackin* to go low i.e. binary 0 and then resets the entire dual-rail data bus to the spacer state. Subsequently in the fourth phase, after an unbounded but a finite and positive time duration, the receiver drives *ackout* low i.e. *ackin* becomes high. One data transaction is now said to have been completed, and the asynchronous circuit stage is ready to process the next data transaction.

The completion detector block [19], which is highlighted in Fig. 1, ensures the complete arrival of all the primary inputs to an asynchronous circuit whether they are valid data or spacer. It consists of an array of 2-input OR gates in the first logic level with each 2-input OR gate used to combine the respective dual-rails of an encoded primary input. The outputs of all the 2-input OR gates are synchronized using a C-element[1] tree.

## 2.2. Asynchronous Function Blocks – Types

Asynchronous function blocks are classified as strongly indicating, weakly indicating, and early output types. Indication implies acknowledging the arrival of the primary inputs to an asynchronous circuit through corresponding monotonic transitions on the intermediate and primary outputs. The transitions are expected to monotonically increase or decrease uniformly throughout the circuit [27]. The general input-output timing characteristics of strong-indication, weak-indication, and early output type asynchronous function blocks are portrayed through Fig. 2.

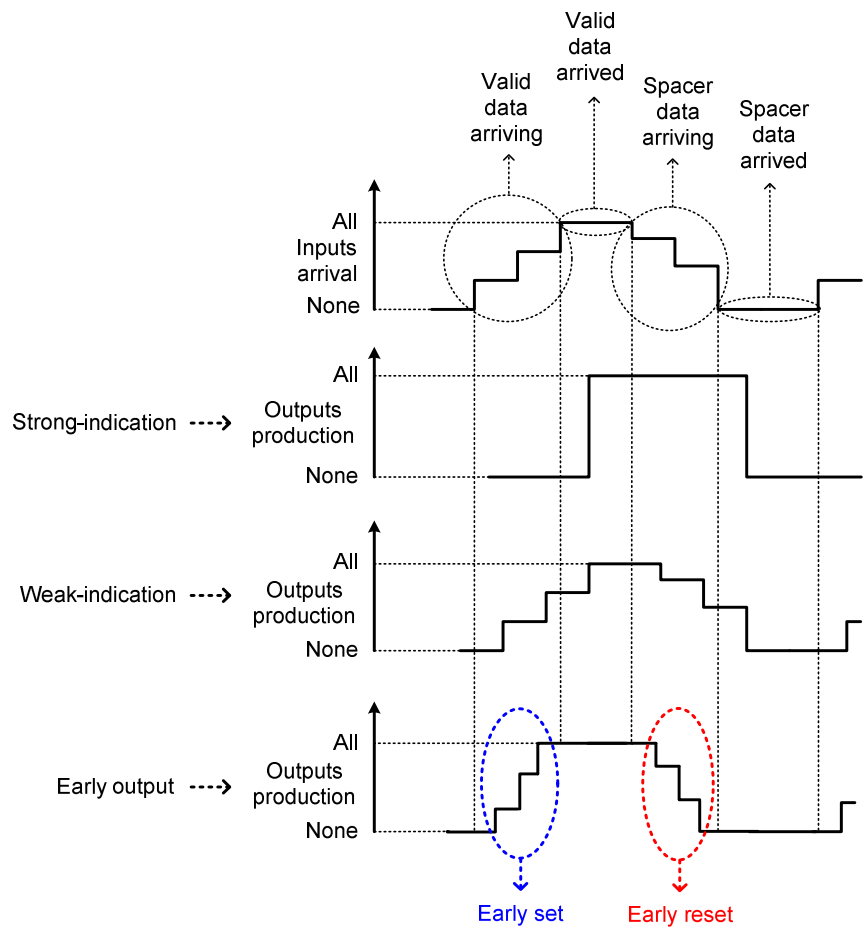

Fig. 2. Depicting inputs-outputs timing correlation of strong-indication, weak-indication, and early output asynchronous function blocks

A strong-indication function block [5] [28] starts data processing only after receiving all the primary inputs, and the required outputs are then produced. A weak-indication function block [5] [29] can commence data processing after receiving just a subset of the primary inputs and can produce some of the primary outputs. However, only after receiving the last primary input, the last corresponding primary output is produced by the weak-indication function block. With respect to weak-indication, the mechanism may be either local or global [30]: local, if the asynchronous function block is internally indicating, and global, if the asynchronous circuit stage provides indication overall. It was shown in [31] that local weak-indication is preferable compared to global weak-indication for robust asynchronous circuit designs. An early output function block [32] [33] is the most relaxed compared to the strong-indication and weak-indication function blocks as it can start data processing after receiving just a subset of the primary inputs and can produce all the primary outputs without having to wait for the arrival of all the primary inputs. In this context, the early output asynchronous function block could exhibit either early set or early reset behavior as shown in Fig. 2. Early set implies that upon receiving a subset of the valid primary inputs, the early output function block produces all the valid primary outputs. The early set property is highlighted through the blue oval in dotted lines in Fig. 2. On the other hand, the early reset behavior implies that upon receiving a subset of the spacer primary inputs, the early output function block processes them and drives all the primary outputs to the spacer state. The early reset property is shown by the red oval in dotted lines in Fig. 2. Relative-timed asynchronous circuits [51] form a subset of the early output circuits, which may incorporate additional timing assumptions to ensure their safe operation.

---

[1] The C-element outputs binary 1 or 0 if all its inputs are binary 1 or 0 respectively. However, if its inputs are different, the C-element retains its existing steady-state output. The 2-input C-element is portrayed by a circle with the marking 'C' in the figures.

## 3. Asynchronous Early Output RCA comprising DBFAs and SBFAs

An asynchronous RCA architecture is used which makes use of DBFAs and SBFAs instead of utilizing only SBFAs. Due to the usage of DBFAs, the number of carry propagation stages within the RCA is almost halved and so the data propagation delay i.e. the (forward) latency would be minimized. An example 32-bit asynchronous RCA that makes use of SBFAs for the least significant positions and DBFAs for the more significant positions is shown in Fig. 3.

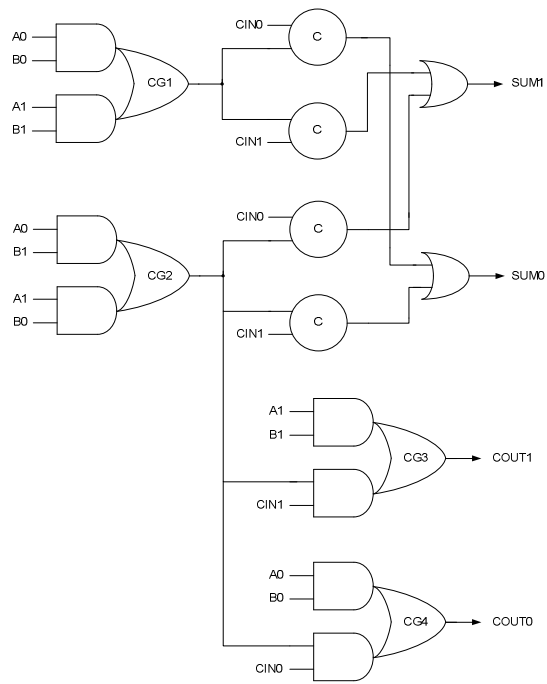

Fig. 4. Early output SBFA [11] with implicit logic redundancy

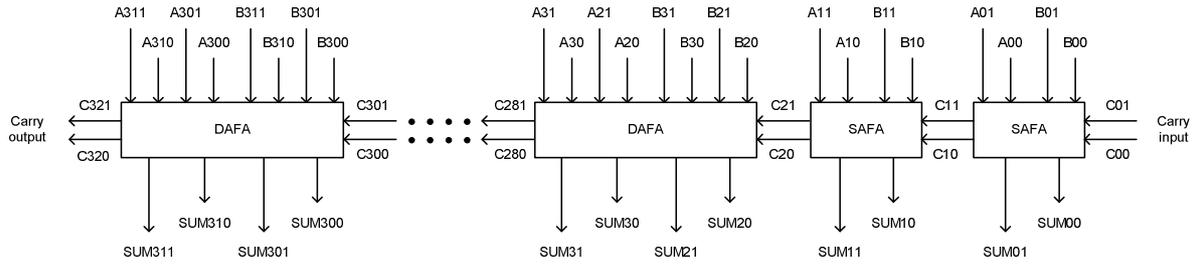

Fig. 3. 32-bit asynchronous RCA composed using SBFAs in the least significant positions and DBFAs in the more significant positions. SBFAs in the least significant adder positions help to reduce the considerable latency that a least significant DBFA is likely to encounter.

$SUM1 = A0B0CIN1 + A0B1CIN0 + A1B0CIN0 + A1B1CIN1$ (1)

$SUM0 = A0B0CIN0 + A0B1CIN1 + A1B0CIN1 + A1B1CIN0$ (2)

$COUT1 = A0B1CIN1 + A1B0CIN1 + A1B1CIN0 + A1B1CIN1$ (3)

$COUT0 = A0B0CIN0 + A0B0CIN1 + A0B1CIN0 + A1B0CIN0$ (4)

Fig. 5. Logic equations corresponding to the dual-rail encoded SBFA. The equations are factorized [42] according to the safe quasi-delay-insensitive (QDI) logic decomposition principles given in [39]. The resulting early output SBFA synthesized is depicted by Fig. 4. For more details, please refer to [11].

$SUM11 = A11A01B10B00CIN0 + A10A01B11B00CIN0 + A11A00B10B01CIN0$
$\qquad + A10A00B11B01CIN0 + A11A00B11B01CIN1 + A11A01B11B00CIN1$
$\qquad + A10A00B10B01CIN1 + A10A01B10B00CIN1 + A10A01B10B01$
$\qquad + A11A00B10B00 + A10A00B11B00 + A11A01B11B01$ (5)

$SUM10 = A11A01B10B00CIN1 + A10A01B11B00CIN1 + A11A00B10B01CIN1$
$\qquad + A10A00B11B01CIN1 + A10A01B10B00CIN0 + A10A00B10B01CIN0$
$\qquad + A11A01B11B00CIN0 + A11A00B11B01CIN0 + A11A00B11B00$
$\qquad + A11A01B10B01 + A10A01B11B01 + A10A00B10B00$ (6)

$SUM01 = A01B00CIN0 + A00B01CIN0 + A00B00CIN1 + A01B01CIN1$ (7)

$SUM00 = A01B01CIN0 + A01B00CIN1 + A00B01CIN1 + A00B00CIN0$ (8)

$COUT21 = A10A00B11B01CIN1 + A11A00B10B01CIN1 + A10A01B11B00CIN1$
$\qquad + A11A01B10B00CIN1 + A10A01B11B01 + A11A01B10B01 + A11B11$ (9)

$COUT20 = A11A01B10B00CIN0 + A10A01B11B00CIN0 + A11A00B10B01CIN0$
$\qquad + A10A00B11B01CIN0 + A11A00B10B00 + A10A00B11B00 + A10B10$ (10)

Fig. 6. Logic equations corresponding to the dual-rail encoded DBFA. The above equations are factorized [42] according to the safe QDI logic decomposition principles discussed in [39]. The resulting early output DBFA synthesized is portrayed by Fig. 7.

Technical NoteAn early output SBFA based on the delay-insensitive dual-rail data encoding is shown in Fig. 4 [11]. (A1, A0), (B1, B0), and (CIN1, CIN0) represent the dual-rail encoded augend, addend, and carry inputs. (SUM1, SUM0) and (COUT1, COUT0) are the dual-rail sum and carry outputs. The early output SBFA shown in Fig. 4 consists of four complex gates (AO22) labelled as CG1 to CG4, four 2-input C-elements (highlighted by the circles with the marking C on their periphery), and two 2-input OR gates. The conjunction of A1 and B1 is visible in the complex gates CG1 and CG3 in Fig. 4. Similarly, the conjunction of A0 and B0 is visible in the complex gates CG1 and CG4. This is an example of implicit logic redundancy [23] [38]. The logic equations corresponding to the SBFA are expressed by (1) to (4) in Fig. 5. Equations (1) to (4) are expressed in the disjoint sum-of-products (DSOP) form [34]. In a DSOP equation, the logical conjunction of any two products yields null [35] [36].

To realize a homogeneously encoded DBFA, the dual-rail code is used to encode the augend, addend, and carry inputs, and the sum and carry outputs. Let (A11, A10) and (A01, A00) represent the dual-rail augend inputs, and let (B11, B10) and (B01, B00) represent the dual-rail addend inputs, and let (CIN1, CIN0) represent the dual-rail carry input. The most significant and least significant dual-rail sum outputs are specified by (SUM11, SUM10) and (SUM01, SUM00) respectively. (COUT21, COUT20) denotes the dual-rail carry output. The logic equations of the DBFA outputs are given by (5) to (10), which are shown in Fig. 6, and the gate level circuit is shown in Fig. 7. Equations (5) to (10) also correspond to the DSOP form.

Fig. 7 shows the new, early output asynchronous DBFA circuit [43]. Fig. 7 contains a mix of discrete gates, complex gates, and 2-input Muller C-elements. However, logic redundancy is implicit in the DBFA design as that of the SBFA design. Both the DBFA and SBFA designs would exhibit early reset behavior during the application of the spacer.

For the homogeneously encoded DBFA shown in Fig. 7, which when positioned in an intermediate position in an RCA, the element present in the critical path of the non-redundant design would be an AO21 complex gate. However, if the DBFA is used for adding the least significant augend and addend inputs bit-pair the critical path traverses through an AO22 gate, a 2-input AND gate, and an AO21 gate. Hence, two SBFAs may be used for the least significant RCA input bit positions to compensate for the delay in the least significant DBFA. In other words, two SBFAs can be used to replace one DBFA to add the least significant input bit pair of the RCA, as shown in Fig. 3.

## 4. Physical Implementation and Design Metrics of Asynchronous RCAs

Many 32-bit asynchronous RCAs [10] [11] [17] [18] [38] [43] were implemented for comparison based on a 32/28nm CMOS technology [37]. The asynchronous RCAs were physically realized by cascading homogeneously or heterogeneously encoded SBFAs and/or DBFAs. The physical circuit architectures of the homogeneously and heterogeneously encoded asynchronous RCAs are given in [38], and the interested reader is referred to the same for details. The 2-input Muller C-element was manually implemented using 12 transistors and it was made available to physically implement the various asynchronous RCAs, registers, and the completion detector. High fan-in C-element functionality where imminent was safely decomposed into 2-input C-elements based on the QDI logic decomposition method [39] which guarantees gate-orphan freedom. The Muller C-element is not made available as part of any commercial standard digital cell library since a commercial digital cell library is primarily meant for synchronous digital circuit design.

An asynchronous circuit stage, as shown in Fig. 1, consists of the asynchronous function block, the input registers, and the completion detector. With respect to the asynchronous function blocks realized based on heterogeneous delay-insensitive data encoding, dual-rail to 1-of-4 encoders would be present before the function block and 1-of-4 to dual-rail decoders would be present after the function block as illustrated in [38]. The input registers and the completion detector would be identical and only the asynchronous function blocks would tend to differ in their physical composition. Hence any differences between the simulation results of the various asynchronous RCAs can be attributed to the differences between their respective adder logic. This paves the way for a direct comparison of the design metrics viz. latency, area, and power dissipation of the different asynchronous RCAs after physical synthesis.

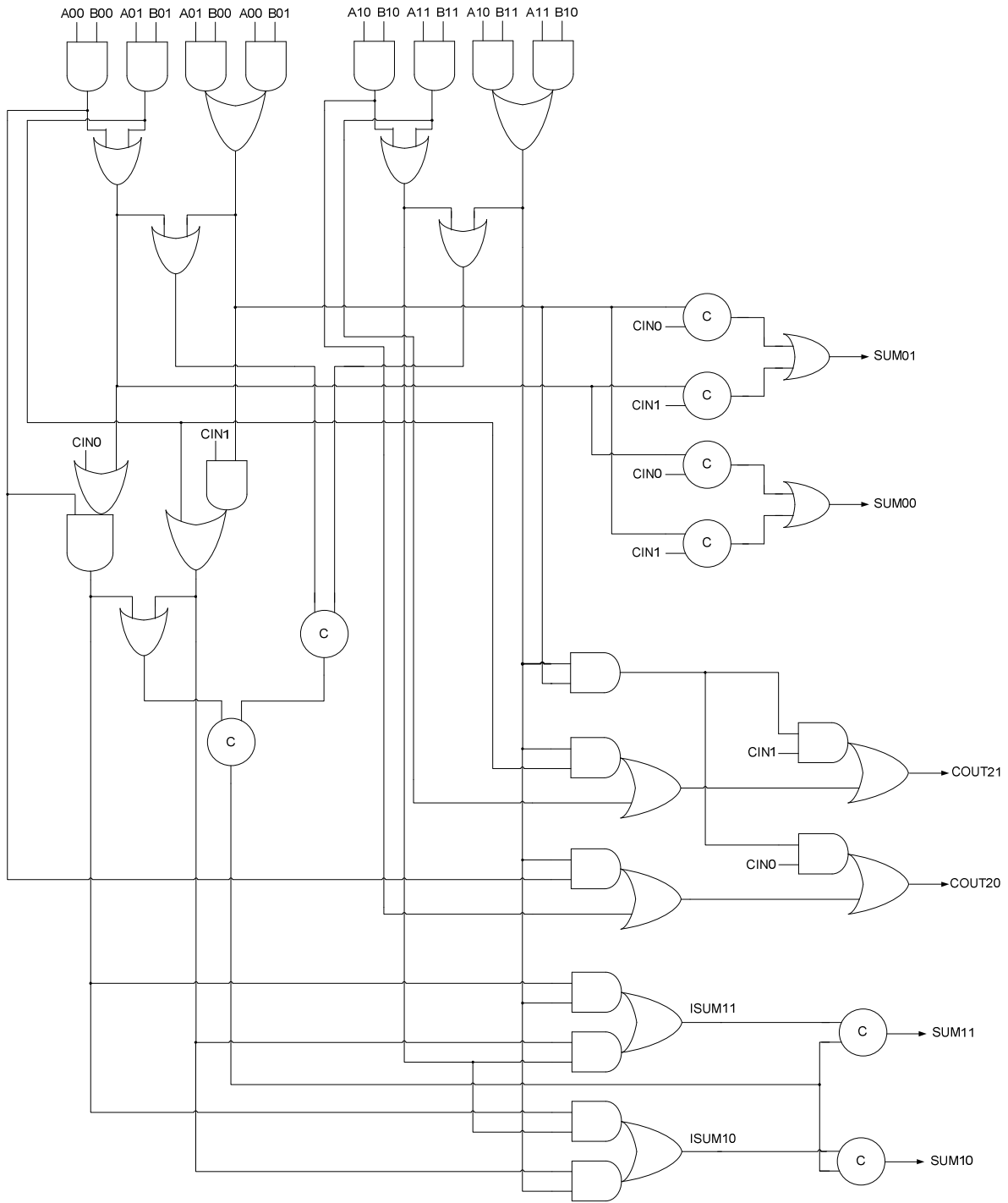

Fig. 7. Proposed area optimized early output DBFA (with implicit logic redundancy)

About 1000 random input vectors were identically supplied to all the asynchronous RCAs at time intervals of 20ns through test benches to verify their functionalities and to capture their respective switching activities. The switching activity files generated through the functional simulations were subsequently used for average power estimation using Synopsys PrimeTime. Since the EDA tool, by default, estimates the critical path timing, the worst-case forward latency was alone estimated for a typical case PVT specification (1.05V, 25ºC) of the standard digital cell library [37]. Appropriate wire load was automatically inserted commensurate with the RCAs while performing the simulations. A virtual clock was used to constrain the input and output ports of the asynchronous RCAs and it did not contribute to the circuit's design metrics.

Table 2 presents the simulation results viz. critical path delay (i.e. forward latency), area occupancy, and average power dissipation of various 32-bit asynchronous RCAs. Adder legends viz. Adder1 to Adder16 are used in Table 2 to simplify the discussion of results. Two general observations can be made from Table 2. Homogeneously encoded asynchronous RCAs feature reduced latency, area, and power dissipation than the heterogeneously encoded asynchronous RCAs. It was shown in [43] that the homogeneously encoded asynchronous RCAs incorporating DBFAs report enhanced optimizations in the design metrics than their heterogeneously encoded RCA counterparts containing DBFAs. While this may be due to the difference in the function blocks, the presence of extra encoders and decoders in the case of the heterogeneously encoded asynchronous circuit causes a disadvantage compared to the homogeneously encoded circuit counterpart. The asynchronous RCA, referred to as Adder14 in Table 2, incorporates only the proposed asynchronous DBFAs. The asynchronous RCA, referred to as Adder15 in Table 2, contains 15 DBFAs (proposed) and 2 SBFAs which reports lesser latency compared to several 32-bit asynchronous RCAs containing only SBFAs. Further, Adder 15 reports lesser silicon area compared to Adder 12. The introduction of more SBFAs into the least significant positions as a replacement for the DBFAs tends to increase the latency although this may reduce the area, as noticed in the case of Adder16 in Table 2. Hence, SBFAs cannot be arbitrarily used to replace the DBFAs to minimize the latency, and the replacement should be guided based on a static timing analysis.

It is noted from Table 2 that with respect to area and power dissipation, Adder1 is optimized than the rest. This is quite expected since the SBFA [11] constituting Adder 1 requires just 27.45$\mu m^2$ of silicon. On the other hand, the DBFA of [43] requires 106.74$\mu m^2$ of silicon, which signifies almost a 3× area increase compared to the area occupancy of the SBFA. The proposed DBFA requires 76.24$\mu m^2$ of silicon. The less area occupancy of the SBFA shown in Fig. 4 translates into a marginally less average power dissipation for Adder1 compared to Adder15 by just 1%. The reduction in power dissipation is rather meagre and this is because all the asynchronous RCAs mentioned in Table 2 tend to satisfy the monotonic cover constraint [19], which implies the activation of a unique signal path from each primary input to a primary output. The monotonic cover constraint is inherent in DSOP expressions, which govern the asynchronous adder outputs.

However, in terms of latency, Adder15 is optimized than Adder1 by 29.4%. It is noted that the proposed early output 32-bit asynchronous RCA, i.e. Adder15 which comprises 2 stages of SBFAs in the least significant positions and 15 stages of DBFAs in the more significant positions achieves the following optimizations in design metrics over its architectural counterparts: i) 18.8% reduction in area than a previously reported 32-bit early output asynchronous RCA which also has 15 stages of DBFAs and 2 stages of SBFAs (Adder 12 [46]), ii) 29.4% reduction in (forward) latency than a 32-bit early output asynchronous RCA containing only SBFAs [11], iii) 33.8% reduction in latency and 33% reduction in area over a weak-indication section-carry based carry lookahead adder (SCBCLA) [44], iv) 28.9% reduction in latency and 30.4% reduction in area over a weak-indication hybrid SCBCLA-RCA [44], v) 20.9% reduction in latency and 23% reduction in area over an early output RCLA [45], vi) 13.8% reduction in latency and 19.4% reduction in area over an early output hybrid RCLA-RCA [45], and (v) a 11% reduction in latency, 34% reduction in area, and 4.8% reduction in average power dissipation over an early output CSLA that features an optimum 8-8-8-8 uniform input partition [41].

Table 2. Design metrics of different 32-bit asynchronous RCAs estimated using a 32/28nm CMOS process

| Adder legend | Adder type and reference(s); Encoding | Forward latency (ns) | Area ($\mu m^2$) | Power ($\mu W$) |
|---|---|---|---|---|
| Adder1 | RCA [11]; Homogeneous, Redundant logic | 3.10 | 1658.80 | 2161 |
| Adder2 | RCA [10]; Heterogeneous, No redundancy | 7.06 | 2016.63 | 2170 |
| Adder3 | RCA [17, 38]; Homogeneous, No redundancy | 4.12 | 2866.49 | 2200 |
| Adder4 | RCA [17, 38]; Homogeneous, Redundant logic | 2.84 | 2931.55 | 2202 |
| Adder5 | RCA [43]; Homogeneous, No redundancy | 4.01 | 2472.06 | 2174 |
| Adder6 | RCA [43]; Homogeneous, Redundant logic | 2.21 | 2488.32 | 2173 |
| Adder7 | RCA [18, 38]; Heterogeneous, No redundancy | 4.36 | 3301.58 | 2191 |
| Adder8 | RCA [18, 38]; Heterogeneous, Redundant logic | 3.03 | 3366.65 | 2192 |
| Adder9 | RCA [43]; Heterogeneous, No redundancy | 4.22 | 2634.71 | 2182 |
| Adder10 | RCA [43]; Heterogeneous, Redundant logic | 2.38 | 2650.98 | 2182 |
| Adder11 | RCA [46]; Homogeneous, Redundant logic; 16 DBFAs | 2.21 | 2488.32 | 2173 |
| Adder12 | RCA [46]; Homogeneous, Redundant logic; 15 DBFAs and 2 SBFAs | 2.14 | 2436.48 | 2173 |
| Adder13 | RCA [46]; Homogeneous, Redundant logic; 14 DBFAs and 4 SBFAs; Early output | 2.21 | 2384.63 | 2171 |
| Adder14 | Proposed RCA; Homogeneous, Redundant logic; 16 DBFAs | 2.19 | 2000.36 | 2183 |
| Adder15 | Proposed RCA; Homogeneous, Redundant logic; 15 DBFAs and 2 SBFAs | 2.19 | 1979.01 | 2182 |
| Adder16 | Proposed RCA; Homogeneous, Redundant logic; 14 DBFAs and 4 SBFAs | 2.27 | 1957.67 | 2180 |

## 5. Conclusions

This work has presented a new area optimized asynchronous early output DBFA based on homogeneous delay-insensitive dual-rail data encoding adhering to the 4-phase return-to-zero handshaking, and a RCA architecture incorporating the asynchronous early output SBFAs in the least significant stages and asynchronous early output DBFAs in the more significant stages. The asynchronous early output RCA that incorporates DBFAs and SBFAs achieves optimization in the latency. The DBFAs and SBFAs incorporate redundant logic. The optimum number of SBFAs and DBFAs to be used for the least significant and more significant adder positions respectively would depend on the adder size. In this work, a 32-bit dual-operand addition was considered, and it was found that using 2 stages of SBFAs in the least significant positions and 15 stages of DBFAs in the more significant positions leads to optimization in the latency over the other architectural counterparts.

In general, homogeneous delay-insensitive data encoding seems to be preferable than heterogeneous delay-insensitive data encoding for a robust asynchronous circuit design. Adder1, mentioned in Table 2, reports the least area and power dissipation while Adder 12 reports the optimum latency. However, in comparison with Adder12, Adder15 achieves significant reduction in area by 18.8% while exhibiting a 2.3% increase in the latency. In the future, the utility of the proposed asynchronous RCA involving DBFAs and SBFAs for the effective realization of multi-operand additions [49] based on the bit-partitioning approach described in [50] could be considered since multi-operand addition operations are predominant in

Technical NoteTechnical Note

digital signal processing applications. Moreover, evaluating the cycle time of the various asynchronous RCAs discussed should be considered since the cycle time determines the rate (speed) at which fresh data can be input to an asynchronous circuit, where the cycle time is the sum of forward and reverse latencies. Also, as a further work, implementing various asynchronous RCAs based on the 4-phase return-to-one handshaking [47] [48] would be considered for an extensive evaluation of their performance. Although an asynchronous DBFA may be useful to achieve less latencies and/or cycle time, an asynchronous SBFA would however remain useful to implement a variety of common arithmetic operations.